\documentclass[%
reprint,
nofootinbib,
amsmath,amssymb,
aps,
prd]{revtex4-2}

\usepackage{bm}
\usepackage{amssymb}
\usepackage{diagbox}

\renewcommand{\d}{\mathrm{d}} 
\newcommand{\vect}[1]{\bm{#1}}
\usepackage{tensor}

\newcommand{\lin}[2]{\tensor[^\varepsilon]{#1}{#2}}
\newcommand{\ep}{\varepsilon}
\newcommand{\Rl}[0]{\lin{R}{}}
\newcommand{\Pl}[1]{\lin{\Phi}{_{#1}}}
\newcommand{\Kerrho}{\varrho}
\newcommand{\KerrDelta}{\triangle}

\begin{document}

\preprint{APS/123-QED}

\title{Linear perturbations of Kerr black hole in quadratic gravity}

\author{{\v S}imon Kno{\v s}ka}
\email{simon.knoska@matfyz.cuni.cz}

\author{David Kofro{\v n}}%
\email{david.kofron@matfyz.cuni.cz}

\author{Robert {\v S}varc}
\email{robert.svarc@matfyz.cuni.cz}
 
\affiliation{
Charles University, Faculty of Mathematics and Physics, Institute of Theoretical Physics,\\
V Hole{\v s}ovi{\v c}k{\' a}ch 2, 180 00 Prague 8, Czech Republic
}

\date{\today}

\begin{abstract}
Employing the Newman--Penrose formalism and following the classic Teukolsky-like approach, we linearise the field equations of quadratic gravity on the Kerr background and combine them with the linearised Ricci and Bianch identities. This leads to constraints on linear perturbations of the Kerr spacetime in quadratic gravity. The resulting differential equations are decoupled in such a way that only the Ricci tensor perturbations need to be found on the Kerr background in order to fully determine the solution. The results are illustrated in the simple non-trivial cases of the Schwarzschild and Minkowski geometries.
\end{abstract}


\maketitle


\section{Introduction}

The idea of modifying General Relativity (GR) has been reflected for a long time, beginning immediately after the formulation of GR \cite{Weyl:1919, Bach:1921} and extending to the present day; see, e.g., \cite{Cliftonetal:2012}. In general, we can conclude that modifications of GR are inevitable either from an observational point of view \cite{Askaretal:2019, PeeblesRatra:2003} or from a purely theoretical perspective \cite{Penrose:1965, GoroffSagnotti:1985}. The latter concern mainly with the ways of extending GR in the strong-field regime. The natural way of modifying GR is to introduce additional higher-curvature terms to the Einstein--Hilbert action. This can be argued as an effective field theory method. Namely, the quadratic curvature terms, whose effects are investigated in this paper, make the theory renormalizable \cite{Stelle:1977} but the price paid is to introduce so-called ghosts in the theory. These ghosts propagate additional degrees of freedom, that is, massive spin-0 and massive spin-2 mediators connected to the presence of up to the fourth-order derivatives. Although troublesome, one can still derive physically relevant consequences below the cut-off energy scale.

The (Stelle) quadratic gravity \cite{Stelle:1978} investigated in this paper contains two prominent subclasses. One of them is $R^{2}$-theory that was of interest in the cosmological models of inflation; see e.g. \cite{Starobinsky:1980}. The second one, Einstein--Weyl gravity, extends the GR action by the square of the Weyl tensor. Interestingly, it was shown that the gravitational effects of the Weyl (spin-2) term can exceed those of the $R^{2}$ (spin-0) in the case of compact stars \cite{BonannoSilveravalle:2021}.

The interesting and well-reasoned settings for studying quadratic gravity implications are black holes. The astrophysically most relevant case is represented by the Kerr spacetime \cite{Kerr:1963}, which describes the geometry of a stationary axially symmetric rotating black hole. This setting is shown to be the only GR stable rotating solution which is astrophysically viable. Thus, there is much focus on finding remnants of strong-field gravity extending GR in the Kerr background, such as gravitational wave analysis \cite{Abbottetal:2017} or black-hole images \cite{EHTetal:2019}. Although available data are in agreement with the Kerr properties, much is unknown with respect to possible tiny discrepancies described by alternative theories. Here, Stelle's quadratic gravity is a natural candidate since the GR vacuum spacetimes themselves also solve the field equations of quadratic gravity and thus provide suitable backgrounds for perturbative analysis. 

Our aim is to follow Teukolsky's approach \cite{Teukolsky:1973}, where the Newman--Penrose (NP) scalar quantities \citep{NewmanPenrose:1962} are perturbed rather than the metric coefficients. In general, this allows one to perform a coordinate-independent analysis of the perturbative modes. The NP formalism permits placing the additional conditions on the perturbations, e.g., their algebraic structure, existence and behaviour of null geodesic congruences, specific regimes of gravitational field coming from geodesic deviation. The NP formulation of generic quadratic gravity has been discussed in \cite{SvarcPravdovaMiskovsky:2023}, therefore here we focus exclusively on the perturbative regime on the Kerr background. An analogous discussion was conducted within $f(R)$ gravity in~\cite{Suvorov:2019}.

The paper is organised as follows. In the remaining part of the Introduction, we summarise the basics of Stelle's quadratic gravity, the NP formalism, and properties of the Kerr background using the Boyer--Lindquist coordinates.  In Section~\ref{sec:EOM}, the main results that describe constraints on the linear perturbations of Kerr geometry implied by quadratic gravity are presented explicitly. In the subsequent Section~\ref{sec:explicit_solutions} we show two simple examples. The useful NP identities are listed in Appendix~\ref{Appendix:Geom}.

\subsection{Quadratic Gravity \label{sec:QTheorofGrav}}

The action of the Stelle quadratic gravity \cite{Stelle:1978}, see also~\cite{Salvio:2018}, reads
\begin{equation}
    \!\! S = \int\! \d x^{4} \sqrt{-g} \left( \frac{1}{\mathsf{k}} R - \mathfrak{a}C_{abcd}C^{abcd} + \mathfrak{b}R^{2} +\mathcal{L}_M\right), \label{eq:S}
\end{equation}
where $\mathsf{k} = 16 \pi$, the coupling constants $\mathfrak{a}$ and $\mathfrak{b}$ are related to the mass of the spin-2 and spin-0 modes, and $\mathcal{L}_M$ describes the matter fields, respectively. The field equations resulting from the action~(\ref{eq:S}) are
\begin{equation}
\begin{aligned}
    &\frac{1}{\mathsf{k}}\left(R_{ab} - \frac{1}{2}R g_{ab} \right) -  4 \mathfrak{a} B_{ab} \\
    &\hspace{2mm} + 2 \mathfrak{b} \left( R_{ab} - \frac{1}{4}Rg_{ab} + g_{ab} \Box -\nabla_{a}\nabla_{b} \right)R = \frac{1}{2}T_{ab} \,, \label{eq:full_field_eqs}
\end{aligned}
\end{equation}
where $B_{ab}$ is the Bach tensor defined as
\begin{equation}
    B_{ab} \equiv \left(\nabla^{c} \nabla^{d} + \frac{1}{2}R^{cd} \right)C_{acbd} \,,
\end{equation}
which satisfies
\begin{equation}
    B_{ab} = B_{ba} \,, \quad B_{ab}g^{ab} = 0 \quad \mathrm{and} \quad \nabla^{b}B_{ab} = 0 \,.
\end{equation}
There is an alternative expression for the Bach tensor,
\begin{align}
	B_{ab} &= \frac{1}{2}\Box R_{ab} - \frac{1}{6}(\nabla_{a}\nabla_{b} + \frac{1}{2}g_{ab}\Box)R + C_{acbd}R^{cd} \nonumber \\
    &+ \frac{1}{3}R R_{ab} - R_{ad}R_{b}^{\;d} + \frac{1}{4}\left(R_{cd}R^{cd} - \frac{1}{3}R^{2} \right)g_{ab} \,,
\end{align}
which can be conveniently used within the discussion of linear perturbations of vacuum Einstein spacetimes with vanishing $\Lambda$ since the last three terms do not contribute.

\subsection{\label{sec:NP_formalism}The Newman--Penrose formalism}

The essence of the NP formalism \cite{NewmanPenrose:1962} is the null tetrad ($\vect{k}$, $\vect{l}$, $\vect{m}$, $\bar{\vect{m}}$), where we assume the normalisation
\begin{equation}
\vect{k}\cdot\vect{l}=-1\,, \qquad \vect{m}\cdot\bar{\vect{m}}=1 \,,
\end{equation}
corresponding to the metric
\begin{equation}
	\vect{g} = - \vect{k}\vect{l} - \vect{l}\vect{k} + \vect{m}\bar{\vect{m}} + \bar{\vect{m}}\vect{m} \,. \label{eq:metric_normalisation}
\end{equation}

Following the standard textbook definition \cite{Stephanietal:2003}, we introduce the Ricci rotation coefficients,
\begin{align}
        \kappa &= - k_{a;b}m^{a}k^{b} \,, & \nu &= l_{a;b}\bar{m}^{a}l^{b}  \,, \label{kappa_nu_def} \\
        \rho &= -k_{a;b}m^{a}\bar{m}^{b}\,, & \mu &= l_{a;b}\bar{m}^{a}{m}^{b}  \,, \\
        \sigma &= - k_{a;b}m^{a}m^{b} \,, & \lambda &= l_{a;b}\bar{m}^{a}\bar{m}^{b} \,, \\
        \tau &= - k_{a;b}m^{a}l^{b} \,, & \pi &= l_{a;b}\bar{m}^{a}k^{b} \,,
\end{align}
and
\begin{align}
        \varepsilon &= \frac{1}{2}(m_{a;b}\bar{m}^{a}k^{b} - k_{a;b}l^{a}k^{b}) \,, \\
        \beta &= \frac{1}{2}(m_{a;b}\bar{m}^{a}m^{b} - k_{a;b}l^{a}m^{b}) \,, \\
        \gamma &= \frac{1}{2}(l_{a;b}k^{a}l^{b} - \bar{m}_{a;b}m^{a}l^{b}) \,, \\
        \alpha &= \frac{1}{2}(l_{a;b}k^{a}\bar{m}^{b} - \bar{m}_{a;b}m^{a}\bar{m}^{b}) \,. \label{alpha_def}
\end{align}
These coefficients are related to the covariant derivatives in the direction of the frame vectors,
\begin{equation}
\begin{aligned}
	D &= k^a \nabla_a \,, & \Delta &= l^a \nabla_a \,, \\
	\delta &= m^a \nabla_a \,, & \bar{\delta} &= \bar{m}^a \nabla_a \,.
\end{aligned}
\label{def_derivatives}
\end{equation}

The projections of the Weyl tensor are defined as
\begin{equation}
\begin{aligned}
    \Psi_{0} &= C_{abcd}k^{a}m^{b}k^{c}m^{d} \,, \\
    \Psi_{1} &= C_{abcd}k^{a}l^{b}k^{c}m^{d}\,, \\
    \Psi_{2} &= C_{abcd}k^{a}m^{b}\bar{m}^{c}l^{d} 
    \,, \\
    \Psi_{3} &= C_{abcd}l^{a}k^{b}l^{c}\bar{m}^{d}\,, \\
    \Psi_{4} &= C_{abcd}l^{a}\bar{m}^{b}l^{c}\bar{m}^{d}\,.
\end{aligned}
\label{def_Weyl_projections}
\end{equation}
\begin{widetext}
The Ricci tensor projections are given by
\begin{equation}
\begin{aligned}
	\Phi_{00} &= \frac{1}{2}R_{ab}k^{a}k^{b} \,, \\
	\Phi_{01} &= \frac{1}{2}R_{ab}k^{a}m^{b} \,, & \Phi_{10} &= \frac{1}{2}R_{ab}k^{a}\bar{m}^{b} \,, \\
	\Phi_{11} &= \frac{1}{4}R_{ab}(k^{a}l^{b} + m^{a}\bar{m}^{b}) \,, & \Phi_{02} &= \frac{1}{2}R_{ab}m^{a}m^{b} \,, & \Phi_{20} &= \frac{1}{2}R_{ab}\bar{m}^{a}\bar{m}^{b} \,, \\
	\Phi_{12} &= \frac{1}{2}R_{ab}l^{a}m^{b} \,, & \Phi_{21} &= \frac{1}{2}R_{ab}l^{a}\bar{m}^{b} \,, \\
	\Phi_{22} &= \frac{1}{2}R_{ab}l^{a}l^{b} \,.
\end{aligned}
\label{def_Ricci_projections}
\end{equation}
It is convenient to define the tetrad components of the Bach tensor,
\begin{equation}
\begin{aligned}
	B_{(0)(0)} &= B_{ab}k^{a}k^{b} \,, & B_{(0)(1)} &= B_{ab}k^{a}{l}^{b} \,,  & B_{(0)(2)} &= B_{ab}k^{a}{m}^{b} \,,  \\
	B_{(1)(1)} &= B_{ab}l^{a}l^{b} \,, & B_{(1)(2)} &= B_{ab}l^{a}m^{b} \,, &  B_{(2)(2)} &= B_{ab}m^{a}m^{b} \,.
\end{aligned}
\label{def_Bach_projections}
\end{equation}
Since the Bach tensor is traceless, it holds ${B_{(2)(3)}=B_{(0)(1)}}$.
\end{widetext}

Finally, in the same way as (\ref{def_Bach_projections}), we denote the energy-momentum tensor projections ${T_{(i)(j)}=T_{ab}\,e^a_{(i)}e^b_{(j)}}$ with $\vect{e}_{(i)}$ representing the null frame vectors.

\subsection{\label{sec:Kerr_BG}Kerr metric}

The famous Kerr metric written in the Boyer--Lindquist coordinates takes the form
\begin{align}
	\d s^{2} &= - \frac{\KerrDelta}{\Sigma}\left(\d t - a \sin^{2}\theta\, \d \phi \right)^{2} + \frac{\Sigma}{\KerrDelta}\d r^{2} + \Sigma\, \d \theta^{2} \nonumber \\
 &\hspace{5mm}+ \frac{\sin^{2}\theta}{\Sigma}\left[ (r^{2}+a^{2}) \d \phi - a\, \d t \right]^{2} \,, \label{Kerr_metric}
\end{align}
where the standard notation is used, namely
\begin{align}
	\KerrDelta &= r^{2} - 2Mr + a^{2} \,, \\
	\Sigma &= r^{2} + a^{2}\cos^{2}\theta \,, \\
	\Kerrho &= r - ia \cos \theta \,. \label{Kerr_metric_functions}
\end{align}

To describe the Kerr geometry within the NP formalism, a proper choice of the null frame is required. The Kinnersley (null) tetrad is a fairly common possibility, namely
\begin{align}
	\vect{k} &= \frac{r^{2} + a^{2}}{\KerrDelta\sqrt{2}}\partial_{t}
        + \frac{1}{\sqrt{2}}\partial_{r}
        + \frac{a}{\KerrDelta \sqrt{2}}\partial_{\phi} \,, \\
	\vect{l} &= \frac{r^{2} + a^{2}}{\Sigma\sqrt{2}}\partial_{t}
        -\frac{\KerrDelta}{\Sigma\sqrt{2}} \partial_{r}
        +\frac{a }{\Sigma\sqrt{2}}\partial_{\phi} \,, \\
	\vect{m} &= \frac{i a \sin \theta}{\bar{\Kerrho}\sqrt{2}} \partial_{t}
        +\frac{1}{\bar{\Kerrho} \sqrt{2}}\partial_{\theta}
        +\frac{i}{\bar{\Kerrho}\sin \theta \sqrt{2}} \partial_{\phi} \,.
\end{align}
Such a frame satisfies the necessary normalisation conditions (\ref{eq:metric_normalisation}). Moreover, it is aligned along principal null directions of the Weyl tensor, which implies
\begin{equation}
    \Psi_{0} = \Psi_{1} = \Psi_{3} =\Psi_{4} = 0 \,,
\end{equation}
manifestly showing the Petrov type D \cite{Kinnersley:1969}. Since the Kerr spacetime solves the vacuum Einstein equations with $\Lambda=0$, the scalar curvature $R$ and the Ricci and Bach tensor projections, (\ref{def_Ricci_projections}) and (\ref{def_Bach_projections}), respectively, vanish identically. The Ricci rotation coefficients of the Kinnersley tetrad are
\begin{equation}
\begin{aligned}
        \kappa &= 0 \,, & \rho &= -\frac{1}{\Kerrho \sqrt{2}}  \,, &  \epsilon &= 0 \,, \\
        \nu &= 0 \,, & \mu &= -\frac{\KerrDelta}{\Sigma \Kerrho \sqrt{2}}  \,, & \beta &= \frac{\cot \theta}{2\bar{\Kerrho} \sqrt{2}} \,, \\
        \sigma &= 0 \,, & \tau &= - \frac{i a \sin \theta}{\Sigma \sqrt{2}} \,, & \gamma &= \mu + \frac{r - M}{\Sigma \sqrt{2}} \,, \\
        \lambda &= 0 \,, & \pi &= \frac{i a \sin \theta}{\Kerrho^{2} \sqrt{2}} \,, & \alpha &= \pi - \bar{\beta} \,,
\end{aligned}
\label{Kerr_rotation_coeff}
\end{equation}
and the only non-zero projection of the Weyl tensor is
\begin{equation}
    \Psi_{2} = - \frac{M}{\Kerrho^{3}}\,.\rule[-1em]{0mm}{1cm}
\end{equation}

\section{Constraints on the linear perturbations of the Kerr geometry\label{sec:EOM}}

Here, we employ the Newman--Penrose-like formulation of the quadratic gravity \cite{SvarcPravdovaMiskovsky:2023} and extend the Teukolsky approach \cite{Teukolsky:1973} to linearly perturb all the NP quantities around the Kerr background, namely
\begin{equation}
    \zeta \rightarrow \zeta + \ep\,\lin{\zeta}{} + \mathcal{O}\left(\ep^{2}\right) \,, \label{generic_perturbation}
\end{equation}
where on the right-hand side of (\ref{generic_perturbation}) $\zeta$ represents the Kerr background quantity while $\lin{\zeta}{}$ is its linear perturbation. This applies not only to all the Ricci spin coefficients and Weyl and Ricci projections but also to the frame vectors; therefore, the derivatives in the frame directions are also perturbed. Then, only terms up to the linear order are preserved and higher orders are neglected in all expressions.

Notice that in terms of the coordinate components, the linear perturbations of the Einstein space with vanishing cosmological constant have to satisfy constraints following from the field equations (\ref{eq:full_field_eqs}), namely
\begin{equation}
\begin{aligned}
&\lin{R}{_{ab}} - \frac{1}{2}\lin{R}{}g_{ab} - 4 \mathfrak{a}\mathsf{k}\lin{B}{_{ab}} \\
&\quad -2 \mathfrak{b}\mathsf{k}\left(\nabla_{a}\nabla_{b} - g_{ab}\Box\right)\lin{R}{} =  \frac{1}{2}\mathsf{k}\,\lin{T}{_{ab}} \,,
\end{aligned}
\label{generic_perturbed_field_eqs}
\end{equation}
where $g_{ab}$ is the background metric, $\nabla_{a}$ is the associated Levi-Civita derivative, $\Box\equiv \nabla^{a}\nabla_{a}$ is the d'Alembert operator, and $\lin{T}{_{ab}}$ represents the matter sources. Moreover, we use the trace of (\ref{generic_perturbed_field_eqs}), namely
\begin{equation}
    (6 \mathfrak{b}\mathsf{k}\,\Box - 1)\lin{R}{}=\frac{1}{2}\mathsf{k}\,\lin{T}{} \,, \label{eq:Trace_EOM}
\end{equation}
with $\lin{T}{}$ being the trace of energy-momentum tensor linear contribution, i.e.,  $\lin{T}{}=\lin{T}{_{ab}}g^{ab}$.

The remaining constraints are obtained by substituting the perturbation definition (\ref{generic_perturbation}) into the quadratic gravity field equations expressed in the Newman--Penrose formalism in \cite {SvarcPravdovaMiskovsky:2023}. Moreover, a specific form of the Kerr background is taken into account, and geometric Ricci and Bianchi identities listed in the Appendix~\ref{Appendix:Geom}, as well as their perturbed versions, are used to eliminate additional degrees of freedom. \emph{Very extensive calculations} yield a set of equations for Ricci tensor perturbations $\lin{\Phi}{_{AB}}$ formulated on the Kerr background, thereby encoding complete information about its disturbances within quadratic gravity. The equations read
\begin{widetext}
\begin{align}
     & \lin{\Phi}{_{00}} - 2 \mathfrak{a}\mathsf{k} \lin{B}{_{(0)(0)}} - \mathfrak{b}\mathsf{k}D D \lin{R}{}=\frac{1}{4}\mathsf{k}\, \lin{T}{_{(0)(0)}}    \,, \label{FEq_per_00} \\
      & \lin{\Phi}{_{11}} -2 \mathfrak{a}\mathsf{k}\lin{B}{_{(0)(1)}} - \frac{1}{2}\mathfrak{b}\mathsf{k}\Big(D\Delta + \delta\bar{\delta}  - \mu D + \bar{\rho} \Delta - \pi \delta - 2 \bar{\alpha} \bar{\delta}\Big)\lin{R}{}=\frac{1}{8}\mathsf{k}\,\left(\lin{T}{_{(0)(1)}}+\lin{T}{_{(2)(3)}}\right) \,, \\
     & \lin{\Phi}{_{01}} - 2 \mathfrak{a}\mathsf{k} \lin{B}{_{(0)(2)}} - \mathfrak{b}\mathsf{k}\Big(D \delta  - \bar{\pi}D \Big)\lin{R}{} =\frac{1}{4}\mathsf{k}\, \lin{T}{_{(0)(2)}}  \,, \\
    & \lin{\Phi}{_{22}} - 2 \mathfrak{a}\mathsf{k} \lin{B}{_{(1)(1)}} - \mathfrak{b}\mathsf{k}\Big( \Delta \Delta + (\gamma + \bar{\gamma})\Delta \Big)\lin{R}{}=\frac{1}{4}\mathsf{k}\, \lin{T}{_{(1)(1)}} \,, \label{FEq_per_11} \\
    & \lin{\Phi}{_{12}} - 2 \mathfrak{a}\mathsf{k} \lin{B}{_{(1)(2)}} - \mathfrak{b}\mathsf{k}\Big( \delta \Delta + \bar{\pi} \Delta - \mu \delta \Big)\lin{R}{}= \frac{1}{4}\mathsf{k}\, \lin{T}{_{(1)(2)}} \,, \\
     & \lin{\Phi}{_{02}} - 2 \mathfrak{a}\mathsf{k} \lin{B}{_{(2)(2)}} - \mathfrak{b}\mathsf{k}\Big(\delta \delta   - (\beta - \bar{\alpha})\delta  \Big)\lin{R}{}=\frac{1}{4}\mathsf{k}\, \lin{T}{_{(2)(2)}} \,, \label{FEq_per_22}
\end{align}
where the Bach tensor components $\lin{B}{_{(i)(j)}}$ are also expressed in terms of the Ricci tensor $\lin{\Phi}{_{AB}}$ perturbations as
\begin{align}
\lin{B}{_{(0)(0)}} &=  \boxdot \lin{\Phi}{_{00}} - \frac{1}{6}D D \lin{R}{}  -4 \left(\bar{\tau} D - \bar{\rho} \bar{\delta} + 2 \bar{\rho} \alpha \right)\lin{\Phi}{_{01}}	-4 \left(\tau D - \rho \delta + 2 \rho \bar{\alpha} \right) \lin{\Phi}{_{10}}	+ 8 \rho \bar{\rho} \lin{\Phi}{_{11}} \,, \label{B_per_00} \\
\lin{B}{_{(0)(1)}} &= \boxdot \lin{\Phi}{_{11}}  - \frac{1}{12}\left(D \Delta + \delta \bar{\delta} - \mu D + \bar{\rho} \Delta - \pi \delta - 2 \bar{\alpha} \bar{\delta} \right)\lin{R}{} +2 \mu \bar{\mu}\lin{\Phi}{_{00}} \nonumber \\
&\hspace{5mm} + 2 \left(\pi \Delta - \mu \bar{\delta} + 2 \alpha \mu - 2 \gamma \pi \right)\lin{\Phi}{_{01}}+ 2 \left(\bar{\pi} \Delta - \bar{\mu} \delta + 2 \bar{\alpha} \bar{\mu} - 2 \bar{\gamma} \bar{\pi} \right) \lin{{\Phi}}{_{10}} + 2 \pi \bar{\tau} \lin{\Phi}{_{02}}+ 2 \bar{\pi} \tau \lin{{\Phi}}{_{20}}  \nonumber \\
&\hspace{5mm} -2 \left(\bar{\tau} D  - \bar{\rho} \bar{\delta}  -2 \bar{\beta} \bar{\rho} \right) \lin{\Phi}{_{12}} - 2 \left(\tau D - \rho \delta  - 2 \beta \rho \right) \lin{{\Phi}}{_{21}}+ 2 \rho \bar{\rho} \,\lin{{\Phi}}{_{22}} \,, \\
\lin{B}{_{(0)(2)}} &= \boxdot \lin{\Phi}{_{01}} -\frac{1}{6}(D \delta - \overline{\pi} D )\lin{R}{}	+2 \left(\bar{\pi} \Delta - \bar{\mu} \delta + 2(\bar{\mu}-\gamma - \bar{\gamma}) \bar{\pi}  \right)\lin{\Phi}{_{00}} \nonumber \\
&\hspace{5mm} + 4 \tau \bar{\pi} \lin{{\Phi}}{_{10}}-2\left(\bar{\tau} D - \bar{\rho} \bar{\delta} + 2 (\alpha - \bar{\beta})\bar{\rho}\right)\lin{\Phi}{_{02}}-4 \left(\tau D - \rho \delta \right)\lin{\Phi}{_{11}}  + 4 \rho \bar{\rho} \lin{\Phi}{_{12}} \,, \\
\lin{B}{_{(1)(1)}} &= \boxdot \lin{\Phi}{_{22}}  - \frac{1}{6} \Big(\Delta \Delta + (\gamma  + \bar{\gamma}) \Delta  \Big) \lin{R}{} \nonumber \\
&\hspace{5mm}+8  \mu \bar{\mu} \lin{\Phi}{_{11}}+ 4 \left(\pi \Delta - \mu \bar{\delta} - 2 \mu \bar{\beta} + 2 \pi \bar{\gamma} \right) \lin{\Phi}{_{12}}+ 4 \left(\bar{\pi} \Delta  - \overline{\mu} \delta - 2 \beta \bar{\mu} + 2 \bar{\pi} \gamma  \right) \lin{{\Phi}}{_{21}} \,, \label{B_per_11} \\
\lin{B}{_{(1)(2)}} &= \boxdot \lin{\Phi}{_{12}} - \frac{1}{6} \Big(\Delta \delta + \tau \Delta - (\gamma  - \bar{\gamma}) \delta\Big)\lin{R}{} + 4  \mu \bar{\mu}\lin{\Phi}{_{01}}+ 2\left(\pi \Delta - \mu \bar{\delta} - 2 \pi (\gamma - \bar{\gamma})  +  2  \mu(\alpha - \bar{\beta}) \right) \lin{\Phi}{_{02}} \nonumber \\
&\hspace{5mm} + 4\left(\bar{\pi} \Delta  -  \bar{\mu} \delta \right) \lin{\Phi}{_{11}} + 4 \tau \bar{\pi} \lin{{\Phi}}{_{21}} - 2\left(\tau D - \rho \delta - 2\rho \bar{\pi}  \right)\lin{\Phi}{_{22}}  \,, \\
\lin{B}{_{(2)(2)}} &= \boxdot \lin{\Phi}{_{02}}  - \frac{1}{6}\Big(\delta \delta  + (\bar{\alpha}-\beta) \delta \Big) \lin{R}{} \nonumber \\
	&\hspace{5mm}
	+4 \left(\bar{\pi} \Delta - \bar{\mu} \delta + 2 \beta \bar{\mu} - 2 \gamma \bar{\pi} \right)\lin{\Phi}{_{01}}
	+ 8 \tau \bar{\pi} \lin{\Phi}{_{11}}
	-4 \left(\tau D - \rho \delta - 2 \rho \bar{\alpha} \right) \lin{\Phi}{_{12}} \,. \label{B_per_22}
\end{align}
Motivated by the Geroch--Held--Penrose (GHP) formalism \cite{GerochHeldPenrose:1973}, the $\boxdot$ operator is introduced as
\begin{equation}
    \boxdot\equiv \Box  - 4\left(\Psi_2+\mu \rho -\pi \tau +c.c.\right) + p\mathcal{V} + q\bar{\mathcal{V}} + \frac{p^{2}}{2}\left(3\Psi_2+4 \alpha\beta + \mu \rho - \pi \tau \right) + \frac{q^{2}}{2}(3\bar{\Psi}_2+4\bar{\alpha}\bar{\beta} + \bar{\mu}\bar{\rho}-\bar{\pi}\bar{\tau}) + 2 pq (\alpha\bar{\alpha} + \beta \bar{\beta}) \,, 
\end{equation}
with $\mathcal{V}$ standing for
\begin{equation}
    \mathcal{V} \equiv 2 \gamma D - 2 \alpha \delta - 2 \beta \bar{\delta}  - \delta \alpha - \bar{\delta} \beta     + \Psi_{2}  + \alpha \bar{\alpha} - \beta \bar{\beta} + 2  \alpha \tau + 2  \beta \bar{\tau} + \pi \tau - \gamma (\rho + \bar{\rho})   \,, \label{boxdot_V_def}
\end{equation}
where the $p$ and $q$ represent the GHP weights of a particular Ricci component on which $\boxdot$ acts, namely
\begin{center}
\begin{tabular}{ccccccc}
 &  $\lin{\Phi}{_{00}}$ &  $\lin{\Phi}{_{01}}$ &  $\lin{\Phi}{_{02}}$ & $\lin{\Phi}{_{11}}$ & $\lin{\Phi}{_{12}}$ & $\lin{\Phi}{_{22}}$ \\ \hline
$p$ &  2 &  2 &  2 & 0 & 0 & -2 \\
$q$ &  2 &  0 &  -2 & 0 & -2 & -2 \\
\end{tabular}
\end{center}
The d'Alembert operator $\Box\equiv \nabla^{a}\nabla_{a}$ acting on a scalar quantity $\psi$ on the Kerr background can be expressed as
\begin{align}
    \Box \psi =&  2\delta\bar{\delta}\psi-2D\Delta\psi-2\mu D\psi +2\bar{\rho}\Delta\psi  +2\pi\delta\psi +4\beta\bar{\delta}\psi \,.
\end{align}
\end{widetext}

The Kerr background quantities to be substituted into the above field equations are listed in subsection~\ref{sec:Kerr_BG}, for example,
\begin{align}
& \mu \rho - \pi \tau = \frac{r^{2} - 2 r M + a^{2} \cos^{2}\theta}{2\Kerrho^{2} \Sigma} =\frac{\Sigma - 2 r M }{2\Kerrho^{2} \Sigma}\,.
\end{align}
Finally, observe that the field equations (\ref{FEq_per_00})--(\ref{FEq_per_22}) with (\ref{B_per_00})--(\ref{B_per_22}) are of massive Klein--Gordon form, and thus convenient parameters can be assigned to the scalar propagation and the Bach (spin-2) propagation, respectively,
\begin{equation}
6 \mathfrak{b}\mathsf{k} = {m_{0}^{-2}} \,, \qquad  2 \mathfrak{a}\mathsf{k} = {m_{2}^{-2}} \,, \label{eq:mass_terms}
\end{equation}
where positivity is assumed; see \cite{Salvio:2018}. Although these constraints on Kerr background perturbations in quadratic gravity take a surprisingly compact form in the NP formalism, their analysis in coordinates and general situations will be complicated.

\section{Explicit solutions \label{sec:explicit_solutions}}
 
Here, as a very simple explicit illustration of the above formalism, we discuss static spherically symmetric perturbations of the Minkowski and Schwarzschild backgrounds, respectively, where both cases belong to the general Kerr class with ${a=0}$. These assumptions lead to a simplified form of the scalar curvature ${\lin{R}{}=\lin{R}{}(r)}$, and the Ricci tensor,
\begin{equation}
\begin{aligned}
	& \lin{\Phi}{_{01}}= \lin{\Phi}{_{02}} = \lin{\Phi}{_{12}}  = 0 \,, \\
	& \lin{\Phi}{_{00}} = \lin{\Phi}{_{00}}(r) \,, \ \lin{\Phi}{_{11}} = \lin{\Phi}{_{11}}(r) \,, \ \lin{\Phi}{_{22}} = \lin{\Phi}{_{22}}(r)\,. \nonumber
\end{aligned}
\end{equation}
In such a case, the \emph{effective} action of the derivative operators on these scalars is only via their radial dependence, namely
\begin{align}
    D = \frac{1}{\sqrt{2}}\frac{\d }{\d r} \,, \ \Delta = -\frac{1}{\sqrt{2}}\left(1-\frac{2M}{r}\right)\frac{\d}{\d r} \,, \ \delta = 0 \,.
\end{align}
Subsequently, the d'Alembert operator reduces to the radial Laplace operator,
\begin{equation}
    \Box = \frac{1}{r^{2}}\frac{\d}{\d r}\left[r^{2}\left(1 - \frac{2M}{r} \right) \frac{\d }{\d r} \right] \,, \label{eq:dAlamber:spherical_sym}
\end{equation}
the non-trivial Bianchi identities (\ref{Bianchi_i}) and (\ref{Bianchi_k}) are
\begin{widetext}
\begin{align}
     (r - 2M)\,\lin{{\Phi^\prime}}{_{00}}- r\,\lin{{\Phi^\prime}}{_{11}}+2\, \lin{\Phi}{_{00}} - 4\, \lin{\Phi}{_{11}}  &= \frac{1}{8}r\, \Rl^\prime \,, \label{eq:Bianchi_Schw_1}\\
    r\, \lin{{\Phi^\prime}}{_{22}} -(r - 2M)\,\lin{{\Phi^\prime}}{_{11}}+2\, \lin{\Phi}{_{22}}- \frac{4}{r}\left(r - 2M\right) \lin{\Phi}{_{11}}  
      &= \frac{1}{8}\left(r-2M\right)\Rl^\prime \,,\label{eq:Bianchi_Schw_2}
\end{align}
where prime denotes the $r$-derivative. The field equations become
\begin{align}
& \left[\Box - m_{2}^{2} - \frac{2}{r^{2}}\left(1 - 2 M\frac{\d}{\d r}\right) \right] \Pl{00} + \frac{4}{r^{2}}\Pl{11}
=  \frac{\mathfrak{m}}{12}\, \frac{\d^{2}}{\d r^{2}}\Rl \,,& \label{eq:EOM:Schw_1}\\
&\left[\Box - m_{2}^{2}- \frac{4 }{r^{2}}\left(1 -\frac{4M}{r} \right) \right] \Pl{11} + \frac{1}{r^{2}}\left(1-\frac{2M}{r}\right)^{2}\Pl{00}  + \frac{1}{r^{2}}\Pl{22} \nonumber \\
& \hspace{45.0mm} = -\frac{\mathfrak{m}}{24}\left[\left(1-\frac{2M}{r}\right)\frac{\d^{2}}{\d r^{2}} - \frac{2}{r}\left(1-\frac{3M}{r}\right)\frac{\d}{\d r} \right]\Rl \,, \label{eq:EOM:Schw_2} \\
& \left[\Box  - m_{2}^{2} - \frac{2}{r^{2}}\left(1 + 2M\frac{\d}{\d r} \right) \right] \Pl{22} + \frac{4}{r^{2}}\left(1-\frac{2M}{r}\right)^2\Pl{11}
= \frac{\mathfrak{m}}{12}\left(1-\frac{2M}{r} \right)^{2}\frac{\d^{2}}{\d r^{2}}\Rl  \label{eq:EOM:Schw_3}\,,
\end{align}
\end{widetext}
where the trace equation (\ref{eq:Trace_EOM}) was used to simplify (\ref{eq:EOM:Schw_2}) and $\mathfrak{m}$ is defined as
\begin{equation}
\mathfrak{m} = 1 - \frac{m_{2}^{2}}{m_{0}^{2}}=1-\frac{3 \mathfrak{b}}{\mathfrak{a}} \,.
\end{equation}

\subsection{Spherically symmetric perturbations on the Minkowski background}

Let us start with the flat background; in addition to ${a=0}$, we set ${M = 0}$ in (\ref{Kerr_metric}) and (\ref{Kerr_metric_functions}). This solution was obtained already in \cite{Stelle:1978} using an explicit coordinate formulation. In our Teukolsky-like approach, the first step is to solve the trace equation (\ref{eq:Trace_EOM}), where the d'Alembert operator (\ref{eq:dAlamber:spherical_sym}) effectively takes the form
\begin{equation}
    \Box = \frac{1}{r^{2}}\frac{\d}{\d r}\left(r^{2}\frac{\d }{\d r} \right) \,. \label{eq:dAlamber:flat}
\end{equation}
The solution of (\ref{eq:Trace_EOM}) then becomes
\begin{equation}
    \lin{R}{}=\frac{1}{r}\left(C_{+}^{0}e^{m_{0}r}+C_{-}^{0}e^{-m_{0}r}\right) \equiv \frac{1}{r}C_{\pm}^{0}e^{\pm m_{0}r} \,, \label{sR_Mink_solution}
\end{equation}
where the notion of masses (\ref{eq:mass_terms}) is used instead of the theory constants, and the plus-minus sign abbreviates the linear combination of specific modes. The non-trivial Bianchi identities (\ref{eq:Bianchi_Schw_1}) and (\ref{eq:Bianchi_Schw_2}) become
\begin{align}
    r\,\lin{\Phi}{}^{\prime}_{00} - r\, \lin{\Phi}{}^{\prime}_{11} + 2\, \lin{\Phi}{_{00}} - 4\, \lin{\Phi}{_{11}} &= \frac{r}{8}\, \Rl^\prime \,, \label{eq:Bianchi_flat_1}\\
    r\, \lin{\Phi}{}^{\prime}_{22} -r\,\lin{\Phi}{}^{\prime}_{11}  + 2\, \lin{\Phi}{_{22}}- 4\,\lin{\Phi}{_{11}}&= \frac{r}{8}\, \Rl^\prime \,, \label{eq:Bianchi_flat_2}
\end{align}
and their sum difference gives
\begin{equation}
r\left(\lin{\Phi}{}_{00}-\lin{\Phi}{}_{22}\right)^{\prime} + 2\left(\lin{\Phi}{}_{00}-\lin{\Phi}{}_{22}\right) = 0 \,. \label{eq:Bianchi_flat_sum}
\end{equation}
The field equations (\ref{eq:EOM:Schw_1})--(\ref{eq:EOM:Schw_3}) imply
\begin{widetext}
\begin{align}
\left[\Box - m_{2}^{2} - \frac{2}{r^{2}}\right] \Pl{00} + \frac{4}{r^{2}}\Pl{11}
&=  \frac{\mathfrak{m}}{12}\, \frac{\d^{2}}{\d r^{2}}\Rl \,, \label{eq:EOM:flat_1}\\
\left[\Box - m_{2}^{2} - \frac{4 }{r^{2}}\right] \Pl{11} + \frac{1}{r^{2}}\left( \Pl{00}   + \Pl{22}\right)   &= -\frac{\mathfrak{m}}{24}\left[\frac{\d^{2}}{\d r^{2}} - \frac{2}{r}\frac{\d}{\d r} \right]\Rl \,, \label{eq:EOM:flat_2} \\
\left[\Box  - m_{2}^{2} - \frac{2}{r^{2}} \right] \Pl{22} + \frac{4}{r^{2}}\,\Pl{11}
&= \frac{\mathfrak{m}}{12}\,\frac{\d^{2}}{\d r^{2}}\Rl   \label{eq:EOM:flat_3}\,,
\end{align}
\end{widetext}
and the difference of (\ref{eq:EOM:flat_1}) and (\ref{eq:EOM:flat_3}) leads to
\begin{equation}
\left[\Box - m_{2}^{2} - \frac{2}{r^{2}}\right]\left(\lin{\Phi}{}_{00}-\lin{\Phi}{}_{22}\right)=0 \,. \label{eq:EOM:flat_1_3}
\end{equation}
Observe that equations (\ref{eq:Bianchi_flat_sum}) and (\ref{eq:EOM:flat_1_3}) can be satisfied simultaneously if and only if
\begin{equation}
    \lin{\Phi}{_{00}} = \lin{\Phi}{_{22}} \,.
\end{equation}
Using this result, the sum of equations (\ref{eq:EOM:flat_1}) and (\ref{eq:EOM:flat_2}) becomes
\begin{equation}
    \left(\frac{\d^{2}}{\d r^{2}} - m^{2}_{2} + \frac{2}{r}\frac{\d}{\d r} \right) \lin{\Phi}{} =
    \frac{\mathfrak{m}}{24}\left(\frac{\d^{2}}{\d r^{2}} + \frac{2}{r}\frac{\d}{\d r} \right) \lin{R}{} \,,
\end{equation}
with $\lin{\Phi}{} = \Pl{00} + \Pl{11}$ and the right-hand side is determined by (\ref{sR_Mink_solution}). The solution then takes the form
\begin{equation}
    \lin{\Phi}{} = \frac{ C_{\pm}^{B}e^{\pm m_{2}r}}{r} + \frac{ C_{\pm}^{0}e^{\pm m_{0}r}}{24 r} \,,
\end{equation}
where $C_{\pm}^{B}$ are new integration parameters while $ C_{\pm}^{0}$ arise from the particular solution. 
Finally, a straightforward substitution of this expression back into the original field equations and the Bianchi identities yields the explicit Ricci components,
\begin{align}
    \lin{\Phi}{_{00}} &= \frac{C_{\pm}^{0}}{12r}\left(1 \mp \frac{2}{m_{0}r}  + \frac{2}{(m_{0}r)^{2}} \right)e^{\pm m_{0}r}\nonumber \\
    &\hspace{5mm} +\frac{C_{\pm}^{B}}{2r}\left(1 \pm \frac{1}{m_{2}r}  - \frac{1}{(m_{2}r)^{2}} \right)e^{\pm m_{2}r}  \,, \\
    \lin{\Phi}{_{11}} &= - \frac{C_{\pm}^{0}}{12r}\left(\frac{1}{2} \mp \frac{2}{m_{0}r}  + \frac{2}{(m_{0}r)^{2}} \right)e^{\pm m_{0}r} \nonumber \\
    &\hspace{5mm}+\frac{C_{\pm}^{B}}{2r}\left(1 \mp \frac{1}{m_{2}r}  + \frac{1}{(m_{2}r)^{2}} \right)e^{\pm m_{2}r} \,,
\end{align}
which corresponds exactly to the solution obtained in \cite{Stelle:1978} using the coordinate approach.

\subsection{Spherically symmetric perturbation on the Schwarzschild background}

For simplicity, let us assume $\lin{R}{} = 0$ (which is a possible solution to the trace equation~(\ref{eq:Trace_EOM}) without sources); however, the Schwarzschild mass parameter remains non-vanishing $M\neq 0$. As in the flat case, the combination of the Bianchi identities (\ref{eq:Bianchi_Schw_1}), (\ref{eq:Bianchi_Schw_2}) and the field equations (\ref{eq:EOM:Schw_1}), (\ref{eq:EOM:Schw_3}) relates $\Pl{00}$ and $\Pl{22}$, namely
\begin{equation}
\Pl{00} = \left(1 - \frac{2 M}{r} \right)^{-2} \Pl{22} \,. \label{Phi_00_Schw}
\end{equation}
Then this relation combined with equations (\ref{eq:Bianchi_Schw_2}), (\ref{eq:EOM:Schw_2}), and (\ref{eq:EOM:Schw_3}) gives
\begin{equation}
\Pl{22} =2\left[1- \frac{4M}{r} +\left(\frac{m_2r}{2}\right)^2 \right]\Pl{11} - \frac{r^2}{2}\,\Box\Pl{11} \,, \label{Phi_22_Schw}
\end{equation}
and
\begin{equation}
\begin{aligned}
&\left[1- \frac{2M}{r} +2\left(\frac{m_2r}{2}\right)^2 \right]\Box\Pl{11} \\
&\quad+\frac{2}{r}\left(1-\frac{2M}{r}\right)\left(1-\frac{3M}{r}\right)\frac{\d}{\d r}\Pl{11} \\
&\quad-2m_2^2\left[1- \frac{4M}{r} +\left(\frac{m_2r}{2}\right)^2 \right]\Pl{11}=0 \,,
\end{aligned}
\label{eq:decP11}
\end{equation}
determining thus the $\Pl{11}$ component that has to be subsequently substituted into the previous expressions for $\Pl{22}$ and for $\Pl{00}$, respectively. 

These constraints are obviously more involved than in the flat case, and it is natural to employ approximative or numerical methods for their analysis. For example, near the regular singular point $r=2M$, corresponding to the background horizon, the curvature components behave like
\begin{align}
\Pl{00} &\approx -\frac{C_\mathrm{div}}{m_2^2(r - 2 M)^{2}} + 2 C_\mathrm{reg}m^2_2 M^{2} \,, \\
\Pl{11} &\approx  C_\mathrm{div}\ln\left(r - 2 M\right) + C_\mathrm{reg}  \,, \\
\Pl{22} &\approx -\frac{C_\mathrm{div}}{4m_2^2M^2} + C_\mathrm{reg} \frac{m_2^2}{2}(r - 2 M)^2 \,,
\end{align}
where only the dominant divergent and regular orders are shown, respectively. Assuming only the regular branch of the solution, i.e., $C_{\mathrm{div}}=0$, this is related to the exact solution presented, e.g., in \cite{LuPerkinsPopeStelle:2015, LuPerkinsPopeStelle:2015b, PodolskySvarcPravdaPravdova:2018, PodolskySvarcPravdaPravdova:2020}. In addition, the divergence is physical in the sense that the projection of curvature $R_{ab}u^au^b$ on the frame of a freely falling observer diverges. Moreover, the discussion of divergences on the horizon is a subtle issue affected by the approximative approach since the existence of the horizon is assumed a priori; however, the generic solution in the full theory is typically horizon-less, see the phase diagram in~\cite{SilveravalleZuccotti:2023}.

Vice versa, the asymptotic behaviour in infinity with $r \to \infty$ becomes
\begin{align}
\Pl{00}\; \approx\; \Pl{11}\; \approx\; \Pl{22} \; \approx \; \frac{e^{\pm m_{2}r}}{r^{1 \mp m_{2} M}} \,.
\end{align}
The connection problem for equation~(\ref{eq:decP11}) is complicated, see numerical simulations in the full theory \cite{LuPerkinsPopeStelle:2015, LuPerkinsPopeStelle:2015b, SilveravalleZuccotti:2023}.

In general, no coordinate or gauge freedom is present in the solution, contrasting to the standard analysis where time-scaling and redefinition of the Schwarzschild mass are allowed. Technically, this means that all the integration constants here are physical degrees of freedom of quadratic gravity only.

\section{Summary}

As our main result, we derive the differential equations (\ref{eq:Trace_EOM}) and (\ref{FEq_per_00})--(\ref{FEq_per_22}) that constrain arbitrary perturbations of the Kerr background governed by a generic quadratic theory of gravity and can serve further in various applications. All perturbations are encoded in terms of the Ricci tensor projections. In general, we employ the Newman--Penrose formalism, which provides the advantage of coordinate-independent statements. Simultaneously, the differential equations are of the second order, and thus their analysis could become easier in comparison with the fourth-order constraints in terms of the metric functions. Moreover, the Ricci tensor projections, and subsequently related Newman--Penrose scalars, can be connected to physically relevant quantities and specific viable assumptions can thus be directly performed with respect to a particular situation to be described, e.g., algebraic structure, behaviour of test observers, spacetime symmetries, etc. Finally, the Kerr background itself, as a vacuum solution to general relativity, does not contribute to the Ricci curvature and thus its any non-trivial part immediately arises as the quadratic gravity correction and can be easily recognised.

As simple examples, the static radial perturbations of the Minkowski and Schwarzschild geometries were presented using this approach. However, a much more interesting and technically challenging discussion of perturbations like, e.g., extension of complete Kerr spacetime within quadratic gravity or effects of matter and dynamics even for simpler backgrounds are postponed for further work.

\begin{acknowledgments}

{\v S.} K. and R. {\v S}. acknowledge the support of the Czech Science Foundation Grant No. GACR 22-14791S and the Charles University Grant Agency
Project No. GAUK 425425. D. K. is grateful for the funding from GACR 23-07457S grant of the Czech Science Foundation. 

\end{acknowledgments}

\appendix

\section{Geometrical identities \label{Appendix:Geom}}

For convenience and to fix the convention, we list the standard identities within the Newman--Penrose formalism, see \cite{Stephanietal:2003}.  Naturally, the Ricci and Bainchi identities are identically satisfied for the Kerr background, however, the perturbed identities bring additional important constraints employed to derive the linearised form of the field equations (\ref{FEq_per_00})--(\ref{FEq_per_22}) with (\ref{B_per_00})--(\ref{B_per_22}) where the Ricci tensor perturbations are the only unknown functions.

\subsection{Commutation relations}

Since the derivative operators $D$, $\Delta$, and $\delta$, see (\ref{def_derivatives}), do not commute even if applied to a function, it is useful to express the commutation relations,
\begin{widetext}
\begin{align}
    \Delta D-D \Delta&=\left(\gamma+\bar{\gamma}\right) D+\left(\epsilon+\bar{\epsilon}\right) \Delta-\left(\bar{\tau}+\pi\right) \delta-\left(\tau+\bar{\pi}\right) \bar{\delta}\,, \label{commut_a} \\
    \delta D-D \delta &=\left(\bar{\alpha}+\beta-\bar{\pi}\right) D+\kappa \Delta-\left(\bar{\rho}+\epsilon-\bar{\epsilon}\right) \delta-\sigma \bar{\delta}\,, \label{commut_b}\\
    \delta \Delta-\Delta \delta &=-\bar{\nu} D+\left(\tau-\bar{\alpha}-\beta\right) \Delta+\left(\mu-\gamma+\bar{\gamma}\right) \delta+\bar{\lambda} \bar{\delta}\,, \label{commut_c} \\
    \bar{\delta} \delta-\delta \bar{\delta} &=\left(\bar{\mu}-\mu\right) D+\left(\bar{\rho}-\rho\right) \Delta+\left(\alpha-\bar{\beta}\right) \delta+\left(\beta-\bar{\alpha}\right) \bar{\delta}\,. \label{commut_d} 
\end{align}

\subsection{Ricci identities}

The Ricci identities connect the spin coefficients (\ref{kappa_nu_def})--(\ref{alpha_def}) with the Weyl (\ref{def_Weyl_projections}) and the Ricci (\ref{def_Ricci_projections}) projections,\footnote{Notice that there is a typo in equation (A11) of \cite{SvarcPravdovaMiskovsky:2023} where $\bar{\beta}$ should be instead of $\beta$, see (\ref{Ricci_g}) here.}
\begingroup
\allowdisplaybreaks
\begin{align}
    D\sigma-\delta \kappa=&\,\sigma\left(3 \epsilon-\bar{\epsilon}+\rho+\bar{\rho}\right)+\kappa\left(\bar{\pi}-\tau-3 \beta-\bar{\alpha}\right)+\Psi_{0}\,,
    \label{Ricci_b} 
     \\
    D\rho-\bar{\delta} \kappa=&\, \rho^{2}+\sigma \bar{\sigma}+\rho\left(\epsilon+\bar{\epsilon}\right) -\bar{\kappa} \tau + \kappa\left(\pi-3 \alpha-\bar{\beta}\right)+\Phi_{00}\,, \label{Ricci_a}\\
    D\tau-\Delta \kappa=&\,\rho\left(\tau+\bar{\pi}\right)+\sigma\left(\bar{\tau}+\pi\right)+\tau\left(\epsilon-\bar{\epsilon}\right) -\kappa\left(3 \gamma+\bar{\gamma}\right) +\Psi_{1}+\Phi_{01}\,,\label{Ricci_c} \\
    D\alpha-\bar{\delta} \epsilon=&\, \alpha\left(\rho+\bar{\epsilon}-2 \epsilon\right)+\beta \bar{\sigma}-\bar{\beta} \epsilon-\kappa \lambda -\bar{\kappa} \gamma+\pi(\epsilon+\rho)+\Phi_{10}\,,\label{Ricci_d} \\
    D\beta-\delta \epsilon=&\, \sigma(\alpha+\pi)+\beta\left(\bar{\rho}-\bar{\epsilon}\right)-\kappa(\mu+\gamma) +\epsilon\left(\bar{\pi}-\bar{\alpha}\right)+\Psi_{1}\,,\label{Ricci_e} \\
    D\gamma-\Delta \epsilon=&\, \alpha\left(\tau+\bar{\pi}\right)+\beta\left(\bar{\tau}+\pi\right)-\gamma\left(\epsilon+\bar{\epsilon}\right) -\epsilon\left(\gamma+\bar{\gamma}\right)+\tau \pi - \nu \kappa+\Psi_{2}+\Phi_{11}-\frac{1}{24}R\,, \label{Ricci_f} \\
    D\lambda-\bar{\delta} \pi=&\, \rho \lambda+\bar{\sigma} \mu +\pi(\pi+\alpha-\bar{\beta})-\nu \bar{\kappa} +\lambda\left(\bar{\epsilon}-3 \epsilon\right)+\Phi_{20}\,, \label{Ricci_g}\\
    D\mu-\delta \pi=&\, \bar{\rho} \mu+\sigma \lambda +\pi\left(\bar{\pi}-\bar{\alpha}+\beta\right) -\mu\left(\epsilon+\bar{\epsilon}\right)-\nu \kappa+\Psi_{2}+\frac{1}{12}R\,, \label{Ricci_h}\\
    D\nu-\Delta \pi=& \,\mu\left(\pi+\bar{\tau}\right)+\lambda\left(\bar{\pi}+\tau\right)+\pi\left(\gamma-\bar{\gamma}\right) -\nu\left(3 \epsilon+\bar{\epsilon}\right) +\Psi_{3}+\Phi_{21}\,,\label{Ricci_i} \\
    \Delta \lambda-\bar{\delta} \nu=&\, \lambda\left(\bar{\gamma}-3 \gamma-\mu-\bar{\mu}\right) +\nu\left(3 \alpha+\bar{\beta}+\pi-\bar{\tau}\right)-\Psi_{4}\,, \label{Ricci_j}\\
    \delta \rho-\bar{\delta} \sigma=&\, \rho\left(\bar{\alpha}+\beta\right)+\sigma\left(\bar{\beta}-3 \alpha\right) +\tau\left(\rho-\bar{\rho}\right) +\kappa\left(\mu-\bar{\mu}\right) -\Psi_{1}+\Phi_{01}\,, \label{Ricci_k}\\
    \delta \alpha-\bar{\delta} \beta=&\, \mu \rho-\lambda \sigma+\alpha \bar{\alpha}+\beta \bar{\beta}-2 \alpha \beta +\gamma\left(\rho-\bar{\rho}\right)+\epsilon\left(\mu-\bar{\mu}\right) -\Psi_{2}+\Phi_{11}+\frac{1}{24}R\,, \label{Ricci_l}\\
    \delta \lambda-\bar{\delta} \mu=&\, \nu \left(\rho-\bar{\rho}\right)+\pi\left(\mu-\bar{\mu}\right)+\mu\left(\alpha+\bar{\beta}\right) +\lambda\left(\bar{\alpha}-3 \beta\right) -\Psi_{3}+\Phi_{21}\,,\label{Ricci_m} \\
    \delta \nu-\Delta \mu=&\, \mu^{2}+\lambda \bar{\lambda}+\mu\left(\gamma+\bar{\gamma}\right)-\bar{\nu} \pi +\nu\left(\tau-3 \beta-\bar{\alpha}\right)+\Phi_{22}\,, \label{Ricci_n} \\
    \delta \gamma-\Delta \beta=&\, \gamma\left(\tau-\bar{\alpha}-\beta\right)+\mu \tau-\sigma \nu-\epsilon \bar{\nu} +\beta\left(\mu-\gamma+\bar{\gamma}\right)+\alpha \bar{\lambda} +\Phi_{12}\,, \label{Ricci_o} \\
    \delta \tau-\Delta \sigma=&\, \mu \sigma+\bar{\lambda} \rho +\tau\left(\tau+\beta-\bar{\alpha}\right) +\sigma\left(\bar{\gamma}-3 \gamma\right)-\kappa \bar{\nu}+\Phi_{02}\,, \label{Ricci_p}\\
    \Delta \rho-\bar{\delta} \tau=&\,-\left(\rho \bar{\mu}+\sigma \lambda\right)+\tau\left(\bar{\beta}-\alpha-\bar{\tau}\right) +\rho\left(\gamma+\bar{\gamma}\right)+\nu \kappa-\Psi_{2} -\frac{1}{12}R\,, \label{Ricci_q}\\
    \Delta \alpha-\bar{\delta} \gamma=&\, \nu(\rho+\epsilon)-\lambda(\tau+\beta)+\alpha\left(\bar{\gamma}-\bar{\mu}\right) +\gamma\left(\bar{\beta}-\bar{\tau}\right)-\Psi_{3}\,. \label{Ricci_r}
\end{align}
\endgroup

\subsection{\label{App:Bianchi}Bianchi identities}

Projections of the Riemann tensor covariant derivative with cyclic permutation of indices give the Bianchi identities
\begingroup
\allowdisplaybreaks
\begin{align}
    0=&-\bar{\delta} \Psi_{0}+D\Psi_{1}+(4 \alpha-\pi) \Psi_{0}-2(2 \rho+\epsilon) \Psi_{1}+3 \kappa \Psi_{2} \nonumber \\
    &- D\Phi_{01}+\delta \Phi_{00}+2\left(\epsilon+\bar{\rho}\right) \Phi_{01}+2 \sigma \Phi_{10}-2 \kappa \Phi_{11}-\bar{\kappa} \Phi_{02} 
    +\left(\bar{\pi}-2 \bar{\alpha}-2 \beta\right) \Phi_{00},\label{Bianchi_a} \\
    0=&+\bar{\delta} \Psi_{1}-D\Psi_{2}-\lambda \Psi_{0}+2(\pi-\alpha) \Psi_{1}+3 \rho \Psi_{2}-2 \kappa \Psi_{3} \nonumber \\
    &+ \bar{\delta} \Phi_{01}-\Delta \Phi_{00}-2\left(\alpha+\bar{\tau}\right) \Phi_{01}+2 \rho \Phi_{11}+\bar{\sigma} \Phi_{02}
    +\left(2 \gamma+2 \bar{\gamma} -\bar{\mu} \right) \Phi_{00}-2 \tau \Phi_{10}- \frac{1}{12}D R,\label{Bianchi_e} \\
    0=&-\bar{\delta} \Psi_{2}+D\Psi_{3}+2 \lambda \Psi_{1}-3 \pi \Psi_{2}+2(\epsilon-\rho) \Psi_{3}+\kappa \Psi_{4} \nonumber \\
    &- D\Phi_{21}+\delta \Phi_{20}+2\left(\bar{\rho}-\epsilon\right) \Phi_{21}-2 \mu \Phi_{10}+2 \pi \Phi_{11}-\bar{\kappa} \Phi_{22} 
    +\left(2 \beta -2 \bar{\alpha} +\bar{\pi} \right) \Phi_{20}- \frac{1}{12}\bar{\delta} R,\label{Bianchi_g} \\
    0=&+\bar{\delta} \Psi_{3}-D\Psi_{4}-3 \lambda \Psi_{2}+2(2 \pi+\alpha) \Psi_{3}+ (\rho -4 \epsilon) \Psi_{4} \nonumber \\
    &- \Delta \Phi_{20}+\bar{\delta} \Phi_{21}+2\left(\alpha-\bar{\tau}\right) \Phi_{21}+2\nu \Phi_{10}+\bar{\sigma} \Phi_{22}-2 \lambda \Phi_{11}
    +\left(2 \bar{\gamma} -2 \gamma -\bar{\mu} \right) \Phi_{20},\label{Bianchi_c} \\
    0=&-\Delta \Psi_{0}+\delta \Psi_{1}+(4 \gamma-\mu) \Psi_{0}-2(2 \tau+\beta) \Psi_{1}+3 \sigma \Psi_{2} \nonumber \\
    &- D\Phi_{02}+\delta \Phi_{01}+2\left(\bar{\pi}-\beta\right) \Phi_{01}-2 \kappa \Phi_{12}-\bar{\lambda} \Phi_{00}+2 \sigma \Phi_{11}
    +\left(\bar{\rho}+2 \epsilon-2 \bar{\epsilon}\right) \Phi_{02},\label{Bianchi_b} \\
    0=&-\Delta \Psi_{1}+\delta \Psi_{2}+\nu\Psi_{0}+2(\gamma-\mu) \Psi_{1}-3 \tau \Psi_{2}+2 \sigma \Psi_{3} \nonumber \\
    &+\Delta \Phi_{01}-\bar{\delta} \Phi_{02}+2\left(\bar{\mu}-\gamma\right) \Phi_{01}-2 \rho \Phi_{12}-\bar{\nu} \Phi_{00}+2 \tau \Phi_{11}
    +\left(\bar{\tau}-2 \bar{\beta}+2 \alpha\right) \Phi_{02}+ \frac{1}{12}\delta R, \label{Bianchi_h}\\
    0=&-\Delta \Psi_{2}+\delta \Psi_{3}+2\nu \Psi_{1}-3 \mu \Psi_{2}+2(\beta-\tau) \Psi_{3}+\sigma \Psi_{4} \nonumber \\
    &- D\Phi_{22}+\delta \Phi_{21}+2\left(\bar{\pi}+\beta\right) \Phi_{21}-2 \mu \Phi_{11}-\bar{\lambda} \Phi_{20}+2 \pi \Phi_{12}
    +\left(\bar{\rho}-2 \epsilon-2 \bar{\epsilon}\right) \Phi_{22}- \frac{1}{12}\Delta R, \label{Bianchi_f}\\
    0=&-\Delta \Psi_{3}+\delta \Psi_{4}+3\nu \Psi_{2}-2(\gamma+2 \mu) \Psi_{3} +(4 \beta-\tau) \Psi_{4} \nonumber \\
    &+\Delta \Phi_{21}-\bar{\delta} \Phi_{22}+2\left(\bar{\mu}+\gamma\right) \Phi_{21}-2\nu \Phi_{11}-\bar{\nu} \Phi_{20}+2 \lambda \Phi_{12}
    +\left(\bar{\tau}-2 \alpha-2 \bar{\beta}\right) \Phi_{22} \,,\label{Bianchi_d}
\end{align}
\endgroup
and as the contraction we get
\begingroup
\allowdisplaybreaks
\begin{align}
   \bar{\delta} \Phi_{01}+\delta \Phi_{10}-D\left(\Phi_{11}+ \frac{R}{8}\right)-\Delta \Phi_{00}
    =&\ \bar{\kappa} \Phi_{12}+\kappa \Phi_{21}+\left(2 \alpha+2 \bar{\tau}-\pi\right) \Phi_{01}+\left(2 \bar{\alpha}+2 \tau-\bar{\pi}\right) \Phi_{10} \nonumber \\
    &-2\left(\rho+\bar{\rho}\right) \Phi_{11}-\bar{\sigma} \Phi_{02}-\sigma \Phi_{20}+\left[\mu+\bar{\mu}-2\left(\gamma+\bar{\gamma}\right)\right] \Phi_{00},\label{Bianchi_i} \\
    \bar{\delta} \Phi_{12}+\delta \Phi_{21}-\Delta\left(\Phi_{11}+ \frac{R}{8}\right)-D \Phi_{22} 
    =&\; -\nu \Phi_{01}-\bar{\nu} \Phi_{10}+\left(\bar{\tau}-2 \bar{\beta}-2 \pi\right) \Phi_{12}+\left(\tau-2 \beta-2 \bar{\pi}\right) \Phi_{21} \nonumber \\
    &+2\left(\mu+\bar{\mu}\right) \Phi_{11}+\left(2 \epsilon+ 2 \bar{\epsilon} -\rho-\bar{\rho} \right) \Phi_{22}+\lambda \Phi_{02}+\bar{\lambda} \Phi_{20}, \label{Bianchi_k}\\
    \delta\left(\Phi_{11}- \frac{R}{8}\right)-D \Phi_{12}-\Delta\Phi_{01}+\bar{\delta} \Phi_{02} 
    =&\ \kappa \Phi_{22}-\bar{\nu} \Phi_{00}+\left(\bar{\tau}-\pi+2 \alpha-2 \bar{\beta}\right) \Phi_{02}-\sigma \Phi_{21}+\bar{\lambda} \Phi_{10} \nonumber \\
    &+2\left(\tau-\bar{\pi}\right) \Phi_{11}+\left( 2 \bar{\epsilon} -2 \rho-\bar{\rho}\right) \Phi_{12}+\left(2 \bar{\mu}+\mu-2 \gamma\right) \Phi_{01}.
    \label{Bianchi_j}
\end{align}
\endgroup
\end{widetext}

\end{document}